# Mining the modular structure of protein interaction networks

Berenstein A.[1,⊗], Piñero J.[2,⊗], Furlong L.I.[2], Chernomoretz A.[1,3*]

[1] Departamento de Física, FCEyN, UBA and IFIBA, CONICET, Pabellón 1, Ciudad Universitaria, 1428 Buenos Aires, Argentina

[2] Research Programme on Biomedical Informatics (GRIB), Hospital del Mar Medical Research Institute (IMIM), Universitat Pompeu Fabra (UPF), C/Dr. Aiguader, 88, 08003 – Barcelona, Spain

[3] Laboratorio de Biología de Sistemas Integrativa, Fundación Instituto Leloir, Buenos Aires, Argentina

[⊗]These authors contributed equally to this work

## Abstract

Cluster-based descriptions of biological networks have received much attention in recent years fostered by accumulated evidence of the existence of meaningful correlations between topological network clusters and biological functional modules. Several well-performing clustering algorithms exist to infer topological network partitions. However, due to respective technical idiosyncrasies they might produce dissimilar modular decompositions of a given network. In this contribution, we aimed to analyze how alternative modular descriptions could condition the outcome of follow-up network biology analysis.

We considered a human protein interaction network and two paradigmatic cluster recognition algorithms, namely: the Clauset-Newman-Moore and the *infomap* procedures. We analyzed at what extent both methodologies yielded different results in terms of granularity and biological congruency. In addition, taking into account Guimera's cartographic role characterization of network nodes, we explored how the adoption of a given clustering methodology impinged on the ability to highlight relevant network meso-scale connectivity patterns.

As a case study we considered a set of aging related proteins, and showed that only the high-resolution modular description provided by *infomap*, could unveil statistically significant associations between them and inter/intra modular cartographic features. Besides reporting novel biological insights that could be gained from the discovered associations, our contribution warns against possible technical concerns that might affect the tools used to mine for interaction patterns in network biology studies. In particular our results suggested that sub-optimal partitions from the strict point of view of their modularity levels might still be worth being analyzed when meso-scale features were to be explored in connection with external source of biological knowledge.

# Introduction

One of the major challenges of systems biology is the understanding of the cellular and molecular basis of high-level biological functionality and complex phenotypes. A promising approach to address these problems relies on the characterization of cellular functionality in terms of a global description of the interwoven set of biochemical reactions that take place inside the cell. This systemic approach has received a lot of attention in recent years, fostered by the ever growing availability of massive amounts of data generated at *omic*-scales.

In this context, the network metaphor has appeared as an appealing framework to organize and unveil global patterns of biological relevance from the deluge of available data. It provides a systematic description language based on pairwise relationships (i.e. network links or edges) between entities of interest (i.e. network nodes or vertices). This approach allowed to uncover the role of connectivity and interaction patterns in the emergence of biological functions [1,2], to assign new functionality to non-annotated gene products [3], to propose biomarkers for several pathologies [4] to gain insights into the genotype-phenotype relationship [5–7], and to establish meaningful associations between pathological phenotypes and disruptive perturbations involving particular regions of the underlying protein interaction networks [5,8–10].

The rationale of the network-based approach is that the analysis of topological features of biological networks can unveil relevant biology. In this context, one recurrent strategy consists on the identification of *central* vertices according to network-based centrality indices, with the hope that meaningful biological entities could be recognized. Following this line of research several seminal studies have suggested, for instance, that hub proteins in the *S. cerevisiae* physical interaction network were more likely to be essential than other proteins, giving rise to the so called centrality-lethality-rule [11–15].

Modular and cluster-based descriptions of biological networks have also received much attention in recent years [16]. In this respect, a lot of effort has been paid to take advantage of meaningful correlations established between topological network clusters (which are formed from nodes which are more densely connected with each other than with their neighborhood), and Hartwell's original idea of "biological functional modules", defined as a group of cellular molecular components and their interactions that carry out a specific biological function [17].

The analysis of the modular structure of molecular biological networks on its own has also drawn a lot of attention as it provides a broad and global description of interaction patterns network to understand the complexity of biological systems. A particular insightful use of network's modular descriptions to unveil their organization was introduced by Guimera and Amaral [18–20]. Taking advantage of the modular organization of the network, and once disjoint network communities were recognized, they proposed to classify network nodes according to their intra and inter-module connectivity patterns into seven different universal roles [18]. To that end, they introduced two observables: the *intra-cluster connectivity, Z,* and the *participation* coefficient, *P* of a node (see Methods for details). While the first parameter describes the degree of a node compared to the degree of nodes that belong to the same community, the second one quantifies to what extent a node connects to different modules. Using this methodology they were able to depict highly informative '*cartographic representations*' of several metabolic networks. Furthermore, they showed that non-hub high-participation nodes, detected in *E. coli* metabolic network, tend to display unusually low evolutionary rates, suggesting that relevant biology could indeed be underpinned with the proposed methodology [19].

Being a module-base scheme, an appealing factor of Guimera's analysis is that it does not rely neither on strictly local (i.e. features involving only properties of a node and its direct neighbors) nor global network features, but rather on connectivity patterns displayed at the meso-scale level. In fact, the very notion of network community is used in order to set the meaningful scale over which the connectivity analysis is performed.

Noticeably, the identification of network modules or communities is in fact a mathematically ill-posed problem, in the sense that there is no such a thing as an *a-priori* objective and hypothesis-free definition of how a *good* cluster should be defined. This results in the co-existence of many different community recognition procedures that might produce different network partitions (see [21] for an extensive review). Moreover, it renders the question of how these methodologies would perform in terms of their ability to unveil biologically significant patterns.

Different network community detection procedures make use of qualitatively different strategies. A wide-spread used family of community detection procedures are based on the optimization of a figure-of-merit known as network modularity [22] whereas other well performing algorithms rely on more information-theoretical considerations [23,24] For instance, according to the *infomap* methodology clusters are defined in order to minimize the average description length of a random walk process taking place over the graph [23].

Each clustering strategy presents its own technical caveats. For example, Fortunato & Barthelemy demonstrated that a theoretical resolution limit exists for modularity-based algorithms. This leads to the systematic merge of small clusters in larger modules, even when the clusters are well defined and loosely connected to each other [25]. Since then, many contributions, mainly developed inside the physic community, further explored this effect, proposed alternative methodologies, and established comparative studies considering ad-hoc benchmark network models [26–31]. In particular, it is now rather well estalished that the modularity function is highly degenerate and that partitions with very different resolutions can have arbitrarily similar modularities [26,32,33]. The *infomap* procedure on the other hand was found to be not severely affected by this resolution limit effect when benchmark networks were considered [30].

Despite these developments, modularity maximization is still one of the most popular techniques for the detection of community structure in graphs. In particular, we found that in consonance with the modularity-based community detection procedure employed by Guimera in its original series of papers, many recent cluster-based analysis of different biological problems were tackled considering slight variations of the same kind of modularity maximization guided algorithms[34–36].

In this context, we aimed to present an analysis to put a word of caution about how the "idiosyncrasy" of the considered algorithms could impact on follow-up biological analysis of real protein interaction datasets. In order to better illustrate our point we concentrated our work on two paradigmatic network community detection procedures: the modularity-based Clauset-Newman-Moore (CNM) methodology [37], and the *infomap* algorithm [23], and focused on two important aspects of the problem. On one hand we explored the associations that could be established between biological functional modules and the identified network communities, analyzing the biological homogeneity of these network structures. On the other, taking into account the Guimera's cartographic role characterization we explored network meso-scale connectivity features induced by the considered network modular descriptions. In connection with this point, we studied the ability of the cluster recognition algorithms to mine connectivity patterns of a protein set of interest in order to detect biologically sensible biases in PIN topological features. In particular, we considered aging related proteins as a case study and investigated whether this complex phenotype could be linked to specific intra/inter modular connectivity pattern.

## Results

### *CNM* and *infomap* mined the PIN modular structure at different resolution levels

The modular organization of the PIN was explored considering the *CNM* and *infomap* procedures. Both methodologies resulted in network partitions displaying similar modularity levels ($Q_{infomap}=0.52$ and $Q_{infomap}=0.54$). These values were much higher than the ones obtained in an ensemble of 1000 randomly rewired versions of the PIN that preserved the original degree distribution ($Q_{infomap-rwn}=0.255\pm0.001$ , $Q_{CNM-rwn}=0.313\pm0.001$) stressing the relevance of second and higher order correlations exhibited by in the real network in connection with the emergence of the observed modular structure.

Although the partitions found by both algorithms attained similar modularity values, large differences were observed in terms of the corresponding community size distributions. For instance, whereas there were no infomap communities exceeding four hundred nodes, the CNM partition included four communities with more than a thousand nodes each.

The number of internal links, $l_{int}$, of a given cluster was a relevant magnitude to understand qualitative features of the obtained partitions, and was used as a proxy of the cluster size (see Fig S2a). In order to visualize how the network nodes were distributed among clusters, we showed in Figure 1 the cumulative cluster-size distribution function, $F_{c-size}$, as a function of $l_{int}$ values. For CNM structures, an abrupt change in $F_{c-size}(l_{int})$ took place for a number of internal links of order $l_{int} \sim \lambda \equiv \sqrt{L}$ (where L is the number of edges of the network). This qualitative change in $F_{c\text{-}size}$ suggested the existence of a dominant size scale in the obtained modular description, as 90% of the total number of network nodes was found inside the 8 largest detected CNM communities displaying $l_{int} > \lambda$ (red filled circles in Fig 1).

On the contrary, for infomap clusters (empty squares in Fig 1), a smooth increase of $F_{c\text{-}size}$ levels was observed. In this case the network mass could be split into network sub-structures which spanned a wide range of cluster sizes and did not present any recognizable natural size scale. The results obtained for the considered PIN agreed with Lancichinetti et al general observations [30]: differently from *CNM*, the *infomap* procedure provided a network modular description of multi-resolution character.

We also found that infomap clusters were virtually included inside CNM modules, as almost 90% of infomap internal links were also internal links in CNM clusters (86% of infomap intra-cluster node pairs were preserved in the alternative CNM partition), and only 66% of CNM internal links were preserved as internal infomap links (5% of the total CNM intra-cluster pair of nodes were preserved under the infomap description). As can be seen from Figure S2b, almost the totality (~99%) of broken CNM-internal links took place in the largest CNM detected structures, and only 1% in CNM clusters of internal-link density values lower than the $\sqrt{L}/2$ level.

Summing-up, all our findings were consistent with a scenario were infomap finer structures were merged into larger assemblies under the CNM description (graphical examples for this general tendency were reported in figure S3). Both partitions reported reconcilable descriptions of the PIN, but the community structure revealed by infomap provided a finer granularity level than the one achieved by the CNM procedure.

## Network structures identified at high resolution levels presented higher biological congruency

We considered the biological homogeneity index, BHI, (see Methods) to investigate to what extent different network structures identified at different resolution levels correlated with external biological evidence. BHI values for the 8 CNM larger communities were depicted as red points in Figure 2 (CNM clusters were ordered according to decreasing size). Green triangles showed the BHI level of the *infomap* partition of clusters included in the respective CNM structure. For each CNM community, boxplots depicted distributions of BHI values estimated for an ensemble of 1000 random shuffling realizations of the corresponding *infomap* labels. The BHI levels of *infomap* partitions were systematically higher than the ones observed for the corresponding *CNM* ones (Figure 2), suggesting that the higher granularity level provided by the first algorithm resulted in a significant increase of the overall biological consistency of the detected structures. We could verify that the gain in functional coherence displayed by *infomap* did not come from cluster-size effects alone, as we found for all cases that more than 95% of the random label reassignments presented lower BHI levels than the value displayed by the original *infomap* partition. These findings supported the idea that *infomap* communities represented meaningful graph substructures with higher levels of biological congruence.

## Functional cartography at different resolutions

Meso-scale topological features of the PIN nodes were analyzed studying how they were distributed over the Z-P plane when the *CNM* and *infomap* procedures were alternatively considered (see Fig 3). Dashed lines in the figure delineated regions corresponding to the seven different universal roles introduced by Guimera [18]. It can be appreciated from both panels, that points were not homogeneously distributed in the plane, but they scattered around three local high-density regions laying on the: ultra-peripheral (Z~ -0.5, P~0), peripheral (Z~ -0.5, P~0.5), and connector (Z~ -0.5, P~0.65) areas. Moreover, the coarser resolution level achieved with the CNM algorithm resulted in a general tendency to assign lower participation coefficient values to network nodes (a more detailed quantification of this effect is provided in Sup Table 1).

This last observation was consistent with the fact that the CNM community detection procedure resulted in larger community structures and consequently presented less intra-cluster surfaces than the infomap methodology. In other words, 'internal' surfaces might appear within large CNM clusters when the infomap partition was considered (Fig S2b), causing a number of originally intra-CNM-cluster links to become edges connecting different infomap clusters.

The implications of these discrepancies are not usually addressed in the network biology literature. In fact, several recent studies in different biological contexts used methodologies based on modularity optimization procedures to characterize PIN nodes in terms of topographic roles [34–36]. For these cases, a low number of high-participation nodes were typically reported. However, we want to stress that this was not an intrinsic network feature. Had the infomap clustering procedure been used for the characterization of those networks, a noticeable increase in the number of high participation role nodes would have been observed (see supplementary figure S4). For instance, Chang *et al* considered two PPI-Yeast Networks to investigate the party-date hub dichotomy using Z-P topological features [36]. The first one was a high-confidence yeast PIN introduced and curated by Batada [13] and presented a giant component of 3801 nodes, and 9742 links. The second considered Yeast PIN was originally proposed by Bertin [38] and had a giant component of 2233 nodes and 5750 links. We analyzed the modular structure of these networks using both, the *CNM* and *infomap* community detection algorithms. We observed that also in the cases, high participation nodes were precluded in the *CNM* description and that the use of the *infomap* clustering procedure resulted in a noticeable increase of this type of nodes. The corresponding Z-P density distributions were included as Supplementary Figure S4.

## Meso-scale connectivity patterns of aging-related proteins

In this section we aimed to investigate to what extent the resolution of the considered modular description could condition the finding of significant and non-trivial correlations between complex high-level phenotypes and PIN's meso-scale connectivity patterns. We focused our attention in a set of gene products related to aging: the aging related genes (ARG). Aging is a complex process associated to several complex diseases, that is affected by both, environmental and genetic factors [39]. A lot of effort has been devoted to characterize the genetic basis of aging and resulted in the identification of genes that: are able to modulate the aging process (e.g. gene mutants that increase maximum lifespan in model organisms or linked to human longevity) [40], display transcriptional changes that correlate with age [41], or show specific DNA methylation patterns [42]. Integrative network-based methodologies have already been employed to provide a system-level understanding of aging [40,43,44]. In particular, Xue *et al* have considered a network model of aging integrating a PPI network with gene expression data [43]. They defined network modules analyzing correlation patterns of gene transcriptional profiles and, in the same spirit of the present contribution, found that aging genes were unevenly distributed in their aging-network. Interestingly, they reported that module interfaces - loosely defined as vertices presenting first neighbors located in different modules- had 2-3 fold enrichment in aging associated genes over that the module's cores. In connection with this last finding, we reasoned that the *participation* feature analyzed in the present contribution is particularly well suited to provide a further quantitative topological description of aging-related genes in the context of PIN analysis.

### Aging genes tended to be at the interfaces of high-resolution clusters

As was already shown in Figure 3 and Table S1, the use of *infomap* gave rise to a noticeable increase in protein vertex *participation* values with respect the *CNM*-based characterization. This effect was particularly evident for ARG nodes. Participation-based ROC curves calculated for ARG genes (AUC$_{infomap}$=0.76 and AUC$_{CNM}$ =0.65) displayed (see Fig S5) statistically significant differences between high and low granularity modular descriptions ($p_v$=2.2 $10^{-16}$, deLong's test). This finding suggested that, when estimated at the finer resolution level provided by *infomap* communities, ARG genes were actually boosted toward much higher relative *participation* levels. Therefore, the *participation* feature estimated using *infomap* cluster's definition could better bring out the same tendency reported in [43] regarding aging-related genes to be located at the interfaces of network-communities.

### Aging genes displayed specific topographical roles

We then explored whether aging-related genes were biased to display specific roles over the network. Results reported in Table1 (ARG-dataset column) showed that *provincial-hub* and *connector-hub* roles exhibited the strongest enrichment in ARG when the *CNM* methodology was adopted. On the other hand *kinless* and *kinless-hub* categories were significantly enriched when *infomap* methodology was considered. Under this last analysis alternative, a 64% larger set of ARG were involved in enriched topographic categories and more extreme significance signal levels were achieved.

### The participation feature highlighted non-trivial connectivity patters of the aging gene set

We further examined whether similar biases could be established from the same interaction data considering other topological features different from the *participation* coefficient (e.g. the degree and the betweenness of a node). This point was particularly relevant in our case, as the considered aging-related gene set happened to display rather high *degree* levels over the PIN (see Fig S6), making the node's *degree* a potential confounding factor for our analysis. In order to de-convolve the *degree* signal from *participation* values we performed a bootstrap analysis for the cartographic role enrichment calculation (see Methods). Due to data scarcity of network nodes of high degree levels we considered for this analysis a reduced aging-related gene set (ARG') obtained discarding the top-10% most connected vertices of the original ARG set (see Methods and Sup.Mat Fig S7 for details). Interestingly, we found that only the *infomap-kinless* category enrichment was significant under the bootstrap analysis (see Table 1, ARG'-dataset column). Hence, these results showed that the high resolution level of the *infomap* community structure allowed highlighting the single non-trivial cartographic role enrichment that could not be explained by the effect of the aging gene-set degree distribution.

### Suboptimal performance of alternative topological descriptors to characterize the aging gene set

We further wanted to examine whether other network features were also able to provide non-trivial evidence to distinguish aging related genes from the rest of the considered protein interaction dataset. In particular, we analyzed the performance of these indicators in connection with their ability to bring out mid/poorly connected ARG genes, i.e. unimportant and non-central nodes from the point of view of their degree level. We thus focused our attention on a subset of PIN by removing the 10% of genes with highest degree values (i.e. removing from the analysis nodes with $k>18$), and examined the use of *infomap*-participation to bring out this subset of ARG genes from PIN data. We compared its performance with: the participation feature estimated at a broader resolution (*CNM*-participation), the node degree, and two alternative measures of a node's information-flow related capabilities: *betweenness* and *bridging* centrality. This last feature quantified to what extent a node was located between well-connected regions (see Material and Methods and [45] for further details).

Figure 4 shows ROC curves obtained for the considered features calculated over the analyzed degree-bounded gene-set. Noticeably, along the false-positive-rate (i.e. 1-specificity) range spanned by infomap-kinless ARG genes (1-specificity values in the interval [0,0.045]) the infomap-participation presented the largest sensitivity among the considered descriptors. In particular, regardless of its absolute performance as a topological predictor, the infomap-participation performed better than the CNM-participation feature (i.e. a participation characterization estimated at a broader resolution), and also better than the degree, betweenness and bridging centralities. This observation agreed with the significant and non-trivial link we found between the infomap-kinless category and this group of genes. Moreover, this result suggested that the infomap-participation feature provided the most effective topological alternative, among the considered ones, to bring out this particular gene-set from protein interaction data (in particular, more effective than other information-flow related quantities like node *betweenness* or *bridging centrality*).

## Discussion

In our analysis of the modular structure of a human PIN we found that both, *CNM* and *infomap* clusterization algorithms, produced high-quality network partitions in terms of achieved modularity levels. However, significant differences arose in terms of the granularity of each description. In particular we verified that the largest structures detected by CNM were further broken up in smaller clusters according to the infomap network modular description.

In concordance with these findings, we observed for the high-granularity network partition a general increase in the number of nodes with high-participation roles. Internal surfaces appeared within large CNM clusters when the infomap partition was considered, causing a number of originally intra-CNM-cluster links to become edges connecting different *infomap* clusters. Importantly, the same behavior was observed when already published data was re-analyzed with the infomap prescription (Fig S4). This finding certainly relativized Guimera's original claim that non-hub kinless nodes were not supposed to be found in real-world networks [18]. We have shown instead that this could eventually arise only as a consequence of the employed community detection methodology, better than reflecting an intrinsic feature of the analyzed network. Furthermore, in our work we found that the observed discrepancies in modular descriptions had non-trivial counterparts in the biological coherence of the detected network structures (*infomap* structures presented greater levels of biological coherence), and secondly in the kind of connectivity patterns each algorithm was able to unveil for the analysis of a considered protein-set of interest.

Studying topological network features of proteins related to aging we observed that they could be significantly linked to low and mid participation hub-roles according to the CNM partition (Table 1). However, should the infomap partition be taken into consideration, the same gene-set would have been found to be significantly enriched in high-participation roles (kinless and kinless-hub categories) instead. Noteworthy, for neither hub category we could rule out that the observed enrichment could have arisen from the particular degree distribution exhibited by the corresponding network nodes, as degree-aware random samples showed associations of similar statistical significance levels than the originally observed one. Only the non-hub high participation kinless role detected within the infomap description, proved to be non-trivially connected to the aging related gene set. This meant that the corresponding association was particularly supported by inter and intra modular connectivity patterns of the network nodes.

Importantly, these results suggested that being associated to high infomap-participation nodes (i.e. nodes mostly located at infomap-cluster's interfaces) these proteins could serve for coordination and/or information flow purposes between modules of specific biological functionality. A paradigmatic example of an infomap-kinless protein related to aging is sirtuin 1. SIRT1 and other members of the sirtuin family (SIRT3 and SIRT6) contribute to healthy aging in mammals [46]. In particular, the association of SIRT1 with aging has been proposed based on its role in several processes such as genomic stability, metabolic efficiency, mitochondrial biogenesis, proteostasis and inflammatory responses related to aging [46]. The proteins encoded by the CDKN2A gene provide another interesting example of aging related gene-products associated to infomap-kinless nodes. This gene gives rise to several isoforms known to function as inhibitors of CDK4 kinase, such as p16 and p19. The levels of both p16 and p19 are correlated to the chronological age of tissues in humans and animal models. More interestingly, the CDKN2A gene locus was found to be associated to several of age-associated diseases in a meta-analysis of GWAS [47]. Based on these evidences, the CDKN2A gene is regarded as the best documented gene that control human aging and is associated to age-related diseases.

A final remark is in order here. Even though we have found that the *CNM* partition presented a slight modularity gain when compared with *infomap*, the higher granularity of the later modular description allowed us to highlight network structures of more biological congruence and statistically significant associations between the cartographic role classification scheme and the analyzed aging related protein-set. A major drawback of the *CNM* partition was the existence of extremely large structures for which no clear associations with Hartwell's functional modules could be established. These findings pointed out that the biological significance of a partition obtained through an optimization procedure of a pure topological figure-of-merit should not be taken for granted. Of course this result did not mean that modularity optimization is a flaw methodology *per-se*. Other modularity optimization heuristics exist apart from the *CNM* procedure that could produce partitions at different resolution levels [21,33,48,49]. However, our results suggested that sub-optimal partitions from the strict point of view of their modularity levels might still be worth being analyzed when meso-scale features were to be explored in connection with external source of biological knowledge.

## Conclusions

In this manuscript we addressed in a systematic manner how two alternative modular descriptions of a biological network could condition the outcome of follow-up network biology analysis. In particular we analyzed the use of two paradigmatic and well-known community recognition algorithms, namely the *CNM* and *infomap* procedures, and thoroughly characterized their performance in terms of the granularity of the corresponding inferred network partitions and the biological homogeneity displayed by the detected network structures.

We observed that the *infomap* partition resulted in a keener description of the network's modular structure than the CNM prescription. Noticeably, we found that *infomap* clusters not only corresponded to congruent structures from the topological perspective, but also displayed higher levels of biological homogeneity. Discrepancies in the cluster resolution level displayed by each algorithm had also impinged on the specific kind of meso-scale connectivity patterns each methodology was able to unveil. In this regard, we presented a thoughtful analysis of differences arising in the significant statistical associations that could be established between intra/inter modular connectivity patterns, and specific protein sets related to complex phenotypes like aging.

In this paper we did not aim to present an exhaustive review of existing clustering procedures, but to raise a word of caution regarding the technical tools usually considered in network biology analysis. At this respect our work illustrates the following apparently trivial but often disregarded consideration: optimal partitions from a strict topological point of view do not always provide the best modular description to highlight biologically relevant patterns. Other features like the resulting partition granularity are worth to be considered as well.

## Material and Methods

We considered the set of protein interactions recapitulated in HIPPIE, an integrated protein interaction network with experiment based quality scores [50]. The high-confidence version of the network (v1.5, downloaded on April 2012) included 31068 interactions among 8277 proteins. We focused our analysis on the giant component of this graph, comprising 8000 nodes and 30835 edges, that we dubbed PIN for future reference. An analysis of several network topological features is included as Supplementary Material (see text ST1 and figure FS1 in Sup. Mat.). In our analysis, we also considered a curated database of genes associated with the human aging phenotype provided by *GenAge* [51]. The downloaded dataset (October 2013) comprised 298 genes, and 261 of them could be mapped to PIN.

### Network topological features

In our work, we took advantage of a handful of topological network features. First, we considered the simplest local node-centrality measure, i.e. the *degree* of a node, defined as the node's number of direct neighbors. A second local centrality measure considered in this work was the *clustering coefficient* of a node [52]. It is defined as the ratio between the actual number of connections between two neighbors and the number of all possible connections of this kind, and it specifies the probability that two randomly selected neighbors of the node of interest were connected to each other, and. In addition, we also made use of the node *betweenness* concept, a global centrality measure defined as the number of shortest paths among all network vertices pairs that traversed across the considered node [53]

The *bridging centrality*, *BC,* is another interesting topological feature devised to explore information-flow related capabilities of a given node [54]. It is a measurement of the extent how well a node is located between well connected regions and it is calculated as the product of two factors: the node's betweenness, *be*, and the bridging coefficient, *bc*, of a node *v*:

$$BC(v) = be(v) * \overbrace{\left(\frac{1/k(v)}{\sum_{i \epsilon \mathcal{N}(v)} 1/k(i)}\right)}^{bc(v)}$$

where *k(v)* is the degree of a node *v* and $\mathcal{N}(v)$ is the direct neighbor sub-graph of node. The *bc(v)* factor then determines the extent of how well the node is located between high degree nodes, assessing the local bridging characteristics in the neighborhood.

## Meso-scale network topological features

Two topological features, introduced by Guimera and Amaral, were central to our analysis: the *intra-cluster connectivity, Z*, and the *participation* coefficient, *P* of a node [19]:

$$Z_i = \frac{k_i - \overline{k_{C_i}}}{\sigma_{k_{C_i}}}, \quad P_i = 1 - \sum_{C=1}^{M} \left(\frac{k_{ic}}{k_i}\right)^2$$

where $k_i$ is the degree of the node-i, $\overline{k_{C_i}}$ is the mean degree of nodes in module $C_i$, $\sigma_{Kci}$ is the standard deviation of the nodes degree in that cluster, $k_{iC}$ is the number of connections of node i to members of cluster C, and M is the total number of communities in the network. These quantities can be considered meso-scale descriptors as they explicitly depend on the network's modular structure. Zi measures how well-connected node-i is with respect to the other nodes in the module. On the other hand, the participation coefficient $P_i$ is close to 1 if node-i links are uniformly distributed among all the modules.

## Network null models

To explore local and global structural properties of PIN we considered two different network null models: an Erdos-Renyi (ER) [55], and a fully rewired version (RW) of the real network [56]. Both control graphs preserved the number of nodes and edges of the original network. The ER graph had the same link density than the original network but, as edges were assigned randomly, it presented no correlations of any order. On the other hand, the RW model preserved the original degree distribution, but lacked second and higher order correlations that might exist in the real graph. All network-related calculations were performed using the R statistical framework (v2.15.1) [57] and the igraph library (v0.6-2) [58].

## Modular network description

We considered two well-established network community recognition methodologies: the Clauset-Newman-Moore (CNM) modularity optimization algorithm [37], and the *infomap* procedure [23] . A brief description of the optimization criteria at use by each algorithm was included as Supplementary Material (see sup. text ST2). A thorough analysis and performance comparison of both algorithms can be found in [27,29,30].

## Biological homogeneity index (BHI) of network partitions

The BHI figure-of-merit measures the degree a given partition embodies biologically meaningful clusters, using a reference set of functional classes [59]. It basically quantifies whether genes placed in the same cluster belong to the same functional class. In our case, we relayed on biological knowledge embedded in Gene Ontology protein annotations. A brief description of the considered BHI was included in the Supplementary Material *ST3*.

## Degree control for role enrichment estimation

A bootstrapping procedure was devised to control the node's degree distribution confounding factor for the role enrichment analysis. For each enrichment test, we considered an ensemble of 1000 control random gene-sets having the same degree distribution than the genes under study. A p-value level was assigned according to the number of random realizations displaying the same or larger effects (over/under representation significance) than the ones observed in the original data. Each random realization was built blindly selecting genes from pools of given degree levels in order to conform the degree distribution displayed by the original gene set (see Sup Material ST4 for details).

## Acknowledgements

Funding: The research leading to these results received support from CONICET (PIP0087), UBACyT (20020110200314), ISCIII-FEDER (PI13/00082 and CP10/00524), the IMI JU under grants agreements n° [115002] (eTOX) and nº [115191] (Open PHACTS)], resources of which are composed of financial contribution from the EU's FP7 (FP7/2007-2013) and EFPIA companies' in kind contribution. The GRIB is a node of the Spanish National Institute of Bioinformatics (INB).

# Tables

**Table 1.** Summary of Fisher statistical association test between the ARG set and cartographic role assignments.

|  |  | *Cartography* |  | *ARG dataset* |  | *ARG' dataset* |  |  |
|---|---|---|---|---|---|---|---|---|
|  |  | *Role* | *N* | *ARG* | *pv* | *ARG'* | *pv* | *pv'* |
| *CNM* | **Non Hubs** | R1 | 3884 | 47 | 1 | 47 | 1 | 1 |
|  |  | R2 | 2964 | 119 | 0.00989 | 119 | 4.60E-05 | 0.488 |
|  |  | R3 | 909 | 47 | 0.00413 | 46 | 0.000382 | 0.121 |
|  |  | R4 | 13 | 1 | 1 | 1 | 0.447 | 0.111 |
|  | **Hubs** | R5 | 58 | 15 | **1.72E-09** | 8 | 0.000584 | 0.417 |
|  |  | R6 | 170 | 32 | **2.19E-15** | 12 | 0.00693 | 0.998 |
|  |  | R7 | 2 | 0 | 1 | 0 | 1 | 1 |
| *Infomap* | **Non Hubs** | R1 | 3162 | 26 | 1 | 26 | 1 | 1 |
|  |  | R2 | 2347 | 51 | 1 | 51 | 1.00E+00 | 1 |
|  |  | R3 | 1610 | 65 | 0.13 | 65 | 0.0558 | 0.286 |
|  |  | R4 | 635 | 67 | **4.95E-18** | 67 | **6.67E-21** | 0 |
|  | **Hubs** | R5 | 12 | 1 | 9.86E-01 | 1 | 0.418 | 0.622 |
|  |  | R6 | 102 | 11 | 2.32E-03 | 7 | 5.07E-02 | 0.98 |
|  |  | R7 | 132 | 40 | **1.00E-27** | 16 | **4.37E-06** | 0.352 |

Summary of Fisher statistical association test between the ARG set and cartographic role assignments considering the *CNM* (first 7 rows) and *infomap* (last 7 rows) modular descriptions are shown in the first 4 columns. Results of the corresponding bootstrap control tests are shown in the last three table columns. The number of network's nodes, ARG nodes, and ARG' nodes, assigned to a given role are displayed in columns: *N*, *ARG*, and *ARG'* respectively. Fdr–adjusted Fisher enrichment p-values are reported in *ARG- pv* and **ARG'-pv** columns, where fdr-adjusted bootstrap p-values (see Methods) are shown in column *pv'*. Cartographic role abbreviation: Ultra peripheral (R1), Peripheral (R2), Connector (R3), Kinless (R4), Provincial Hubs (R5), Connector Hubs (R6), Kinless Hubs (R7).

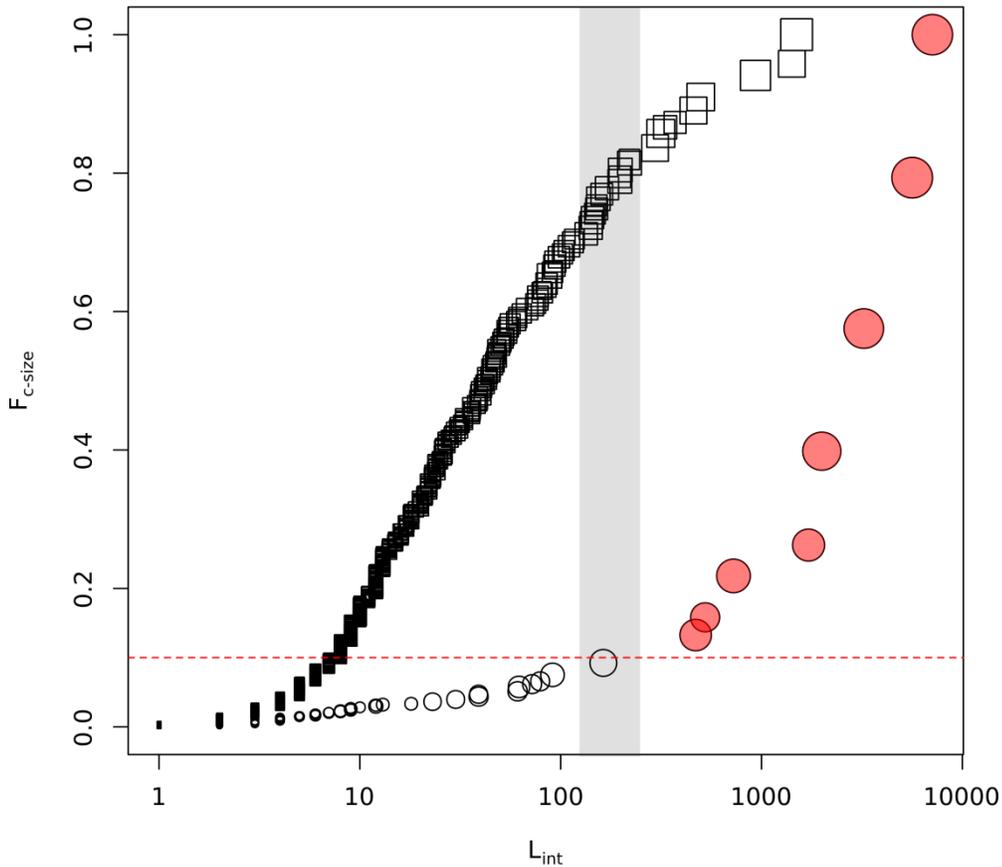

Figure 1-. Analysis of cluster partitions obtained with *CNM* (circles) and *infomap* (empty squares) methodologies. The panel shows the cumulative cluster-size function as a function of $l_{int}$. Symbol sizes were set using a scale proportional to the log-size of the corresponding cluster. The horizontal line corresponds to the 10% accumulated mass level. Dashed vertical lines delimit a region of values of the order square root of total number of links in the network, $[\sqrt{L/2}, \sqrt{2L}]$§, corresponding to the natural scale found to operate in modularity optimization procedures [Fortunato2007, Lancichinetti2011].

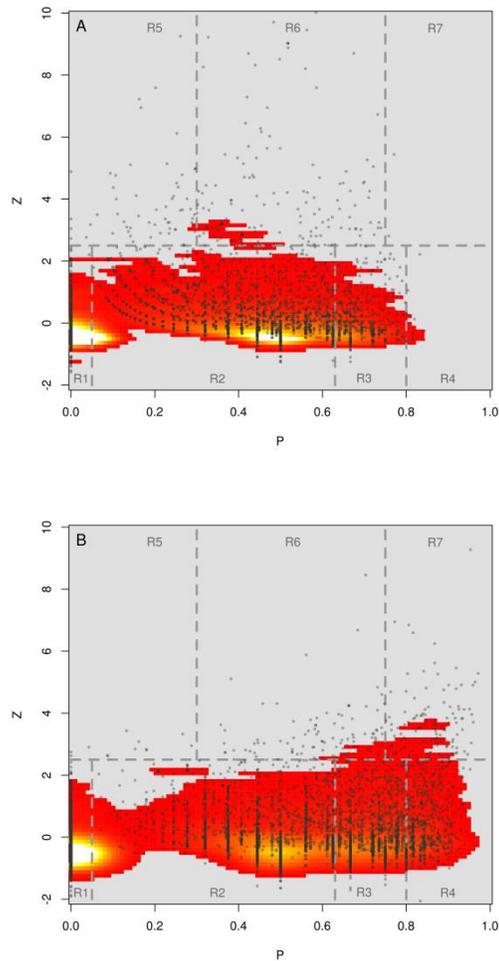

Figure 2 - The Biological Homogeneity Index, *BHI*, estimated for each of the 8 CNM largest communities, is depicted as red points (CNM clusters are ordered according decreasing size). Green triangles show the BHI level of the *infomap* partition of clusters included in the respective *CNM* structure. For each CNM community, boxplots depict main features of BHI distributions estimated for an ensemble of 1000 random shuffling realizations of the corresponding infomap labels. Noticeably, the mean BHI values of the randomized partitions agree

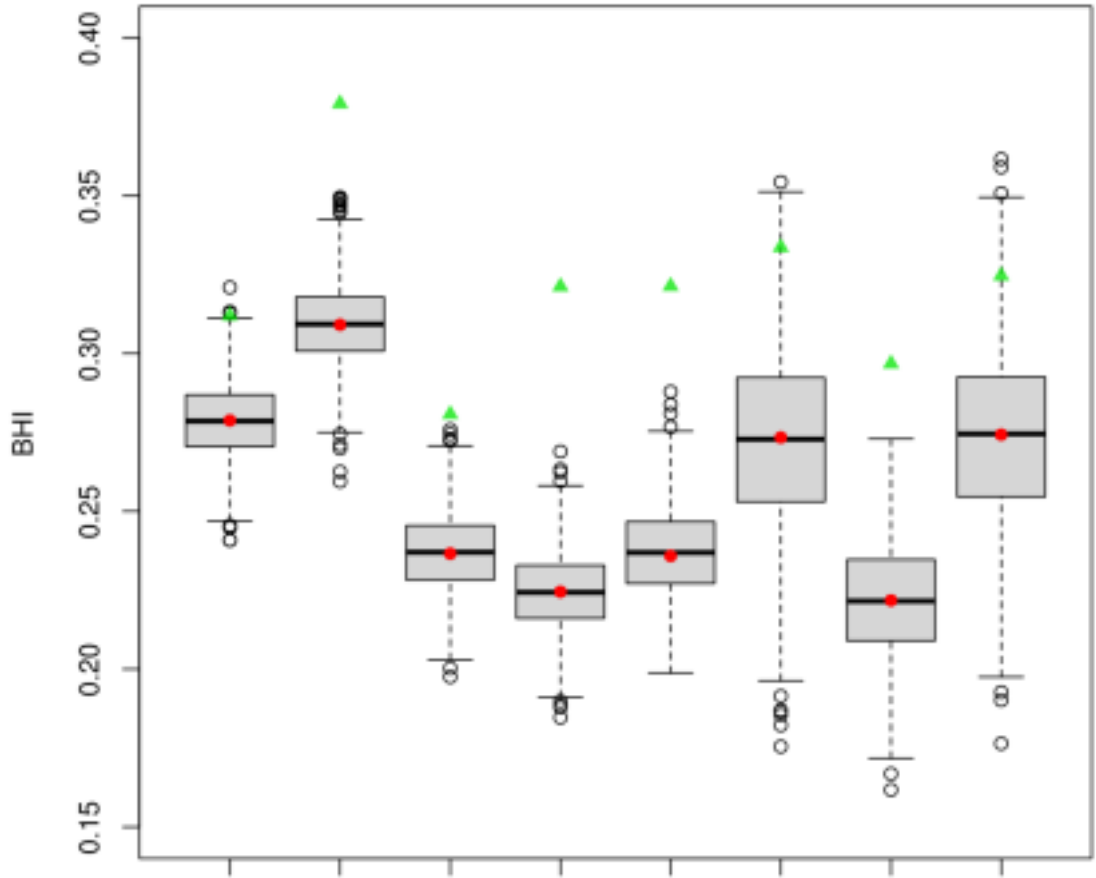

Figure 3 - Distributions of *PIN* nodes in the *Z-P* plane obtained when the *CNM* and *infomap* clusterization were considered are shown in top and bottom panels respectively. A color-coded kernel density estimation was also depicted in the figures. Dashed lines in the figure delineate regions corresponding to the seven different universal roles [6]

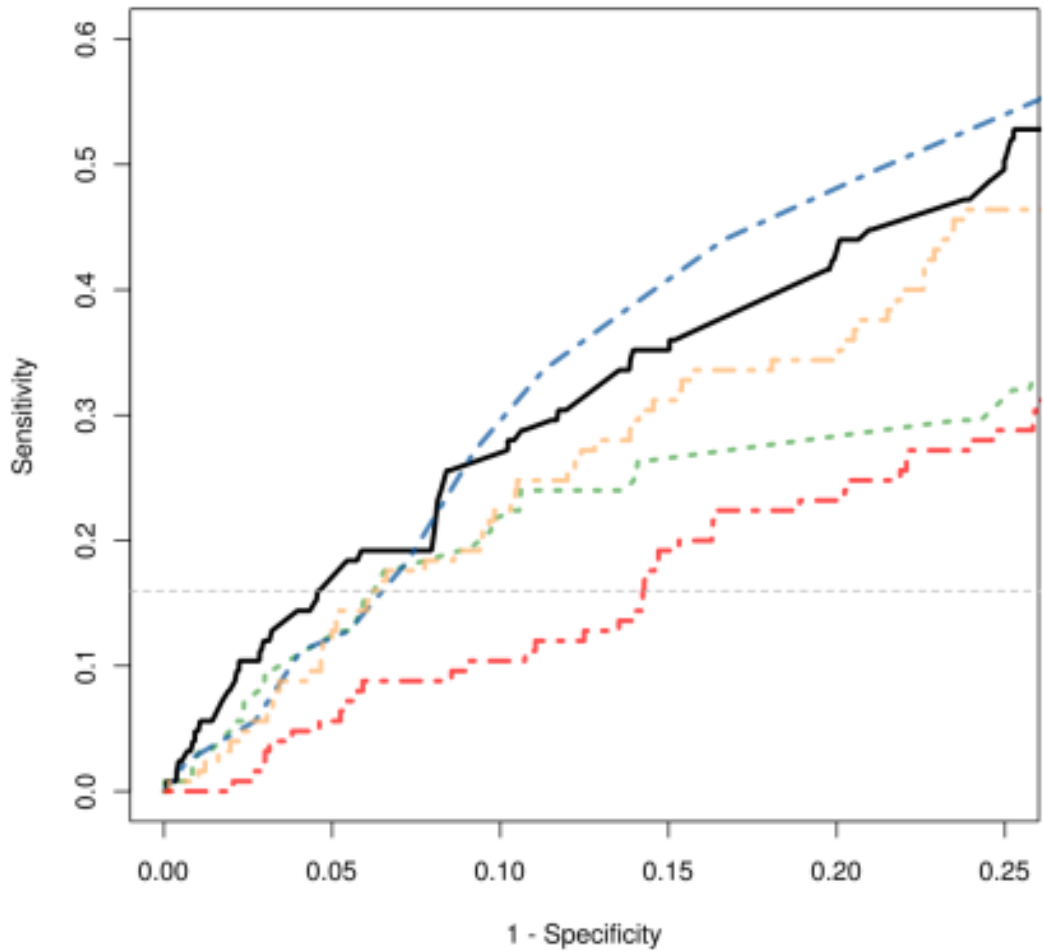

Figure 4 - ROC curves for ARG genes based on node's *infomap*-participation, *CNM*-participation, degree, bridge-centrality and betweenness feature levels are shown as solid black, orange dashed, blue dotted-dashed, dotted-dashed red and green dashed lines respectively. Only nodes with mid/low connectivities (i.e. network nodes with degree values lower than the 90% percentile of the entire degree distribution) were considered. The horizontal dotted line depicts the maximum sensitivity level achieved *infomap-kinless* ARG gene with the lowest *infomap*-participation value

# Supplementary Material

## Mining the modular structure of protein interaction networks

Berenstein A.*[1], Piñero J.*[2], Furlong L.[2]., Chernomoretz A. [1,3]

### ST1 - Protein interaction network characterization

We began our analysis summarizing main general topological features related to the *degree* distribution, *clustering coefficient* [1], and *betweenness* [2] of the nodes of the considered HN. These quantities asses for the number of neighbors, the connectivity among them, and the node relevance in terms of global information flux over the entire network, respectively.

A network characterization in terms of node degree distribution involved one of the most basic and intuitive connectivity-related notion of centrality. The protein interaction network exhibited a heavy-tailed empirical degree distribution - gray points in Figure S1 - reflecting large fluctuations in vertex connectivity (while the network nodes presented an average degree of $\acute{k} = 7.7$, the maximum degree was $k_{max}=492$).This feature is not recapitulated in the ERN null model (see the Method section for a brief description of the considered network null models) - yellow points in Figure S1a - for which the existence of a natural scale for vertex connectivity can be recognized. This fundamental discrepancy has already been widely reported for many different real world networks, and highlights the existence of non-trivial correlation patterns at the level of vertices connectivity.

The heterogeneity observed in neighbor numbers implies the existence of unusually highly connected hub nodes that could act as general shortcuts, globally shortening geodesic distances over the entire network. However, in terms of global information spreading capabilities, additional non-trivial relevant network nodes might also exist. The *betweenness* centrality concept aims to exploit this information-flux point of view to characterize structural properties of network nodes. Figure S1b displays the node's *betweenness* as a function of the node's degree for HN, ERN, and RWN (shown as gray, yellow and red circles respectively). A monotonic and increasing relationship can be recognized for all the three considered networks, denoting the somehow expected general positive correlation trend that exists between the degree and betweenness of a node. Noticeably, for a given degree value, the bio-molecular network presented a wider distribution of betweenness levels. In addition, it presented a higher fraction of high-betweenness and low-degree nodes than the two alternative random network models. If only low connectivity vertices (e.g. having less $\acute{k} = 7.7$ neighbors) were considered, forty nine nodes would be found within the top-10% betweenness score ranking in HN, while four and none in the RWN and ERN cases respectively. These *bottleneck* nodes, that are overrepresented in the real network, constitute an *a priori* interesting subset of proteins since they could play a central intermediary role in information transmission processes taking place over the entire network [2,3] .

Insights about local connectivity patterns can also be gained looking at the node's *clustering coefficient* that quantifies the connectivity among first neighbors of a given node. Figure S1cdisplays the *clustering coefficient* as a function of the node's degree for HN, ERN, and RWN (shown as gray, yellow and red circles respectively). An expected general negative correlation trend can be observed between these two quantities for the three networks. However a much larger number of nodes presenting high clustering coefficient values can be observed for HN.

The presented results suggested the existence of non-trivial topological heterogeneities compatible with a putative underlying modular organization in HN, our bio-molecular network of interest, similarly to what have already been reported in other biological inspired network-based analysis [4-6]. Moreover, the observed structural differences with respect to randomized networks allow us to anticipate that HN interconnectivity patterns probed by different topological observables could highlight non-trivial network components, such us high-betweenness, low-connectivity proteins that might act as important links between modular structures.

### *ST2 - Clustering procedures*

The *CNM* algorithm looks for communities by direct optimization of the modularity *Q*of the graph, that is defined, up to a multiplicative constant, as the number of edges falling within groups minus the expected number in an equivalent network with edges placed at random:

$$Q = \frac{1}{2L} \sum_{ij}^{N} \left( A_{ij} - \frac{k_i k_j}{2L} \right) \delta(C_i, C_j) \qquad [1]$$

In the above equation, $k_i$ is the degree of node-*i*, *L* is the total number of network edges, *N* the total number of nodes, $A_{ij}$ is the adjacency matrix of the network ($A_{ij}=1$ if there is a link between nodes *i* and *j*, and zero otherwise). $C_i$identifies the cluster that includes node-*i*, and$\delta(C_i, C_j)$is a delta function (i.e. $\delta(C_i, C_j) = 1$if node-i and node-j belong to the same cluster, and zero otherwise).

On the other hand, the *infomap* algorithm relies on very different optimization criteria. Clusters are defined in order to minimize the average description length of random walk process taking place over the graph. A two level hierarchy, involving a community tag and a within-community ID tag,is used to identify each network node. As random walkers are expected to expend a lot of time inside densely structures in the graph (i.e. communities), the algorithm iteratively search for node-tagging schemes that produce increasingly compact descriptions of the random walk process. As a by product,a sensible description of the network modular structure is achieved. The *infomap* objective function can be thought in terms of the entropy associated to the random walk process and involved two contribution terms. The first one represents the entropy of the movement between modules, while the second one corresponds to movements within modules:

$$L(P) = q_{inter}H(Q) + \sum_{i}^{s} p_{intra}^{i} H(P^{i}) \qquad [2]$$

Here, $q_{inter}$ is the probability that the walker switches clusters, $H(Q)$ is the entropy associated to between clusters transitions, $p_{intra}^{i}$ is the fraction of movements occurring inside cluster $i$, and $H(P^{i})$ is the entropy of movements within the cluster $i$ .[7].

As can be seen from equations (1) and (2) both considered algorithms rely on very different assumptions and optimization criteria, and thus could provide in principle alternative and complementary descriptions of the modular structure of the analyzed network. A more detailed and general description of optimization criteria and performance comparison of both algorithms can be found in [8,9].

### *ST3 – BHI*

Following Datta&Datta[10], we considered a partition of $k$ clusters, $\{D_1, D_2, ..., D_k\}$, and assumed that $C(x)$ is a functional class containing gene $x$. The biological homogeneity index of the partition resulted:

$$BHI = \frac{1}{k}\sum_{j=1}^{k} \frac{1}{n_j(n_j - 1)} \sum_{x \neq y \in D_j} I(C(x) = C(y))$$

where $n_j$ is the size of cluster-$j$. The indicator function $I(C(x) = C(y))$ takes the value 1 if $C(x)$ and $C(y)$ match. We made use of the functionality implemented in the *clValid* R-package[11], and disregarded functional GO classes annotated under the IEA (inferred from electronic annotations) evidence code.

### *ST4-Degree-aware bootstrap for topographic role enrichment*

A bootstrapping procedure was devised for the topographic role enrichment analysis of the considered gene-groups in order to control for the node's degree distribution factor. For each enrichment test, we considered an ensemble of 1000 control random gene-sets having the same degree distribution than genes under study, and a p-value level was estimated according to the number of random realizations displaying the same or larger effects (over/under representation significance) than the ones observed in the original data.

Each random realization was conformed according the degree displayed by the original gene set randomly extracting genes from pools of given degree levels. In order to warrant for a non-biased sampling, we binned by degree the available sampling nodes requiring a minimal sample size of 100 nodes per bin. Once the non-uniform binning was established we made an *a-posteriori* analysis to make sure that the degree distributions of control random realizations had similar statistical features than the observed one. To that end, we identified which quantile level of the original data was not duly sampled looking for cases where the corresponding observed degree were not included in the inter-quartile range of the respective control realizations. For instance, it can be appreciated form figure S7 that the high degree level of the top 10% of the ARG set, could not be reproduced by the random sampling procedure. In this case the bootstrap analysis was performed considering a reduced ARG' set, discarding the 10% most connected aging related genes.

|         |       | CNM  |      |     |     |     |     |     |       |
|---------|-------|------|------|-----|-----|-----|-----|-----|-------|
|         |       | R1   | R2   | R3  | R4  | R5  | R6  | R7  | Total |
| Infomap | R1    | 3009 | 152  | 1   | 0   | 0   | 0   | 0   | 3162  |
|         | R2    | 716  | 1497 | 84  | 0   | 18  | 32  | 0   | 2347  |
|         | R3    | 146  | 936  | 500 | 0   | 7   | 21  | 0   | 1610  |
|         | R4    | 9    | 304  | 288 | 11  | 4   | 19  | 0   | 635   |
|         | R5    | 3    | 6    | 0   | 0   | 3   | 0   | 0   | 12    |
|         | R6    | 1    | 44   | 5   | 0   | 18  | 34  | 0   | 102   |
|         | R7    | 0    | 25   | 31  | 2   | 8   | 64  | 2   | 132   |
|         | Total | 3884 | 2964 | 909 | 13  | 58  | 170 | 2   | 8000  |

**Sup Table 1.** Distribution of cartographic role assignments according to the Infomap and CNM descriptions. Cartographic role abbreviation: Ultra peripheral (R1), Peripheral (R2), Connector (R3), Kinless (R4), Provincial Hubs (R5), Connector Hubs (R6), Kinless Hubs (R7). The coarser resolution level achieved by the *CNM* algorithm resulted in a general tendency to assign lower participation coefficient values to network nodes. For instance 68% of *infomap*-connector vertices were assigned to lower participation roles (59% peripheral, 9% ultra-peripheral) when the CNM procedure was considered. More strikingly, the majority (94%) of the 635 *infomap*-kinless nodes were re-classified as: CNM-connectors (45%), CNM-peripheral (48%), and CNM-ultra-peripheral nodes (1%). Finally, it can also be observed that nodes originally assigned to hub-like roles when *infomap* procedure was employed were not only affected by this lowering effect in the participation, but in addition, almost 50% of them were also reassigned to non-hub roles when CNM was considered

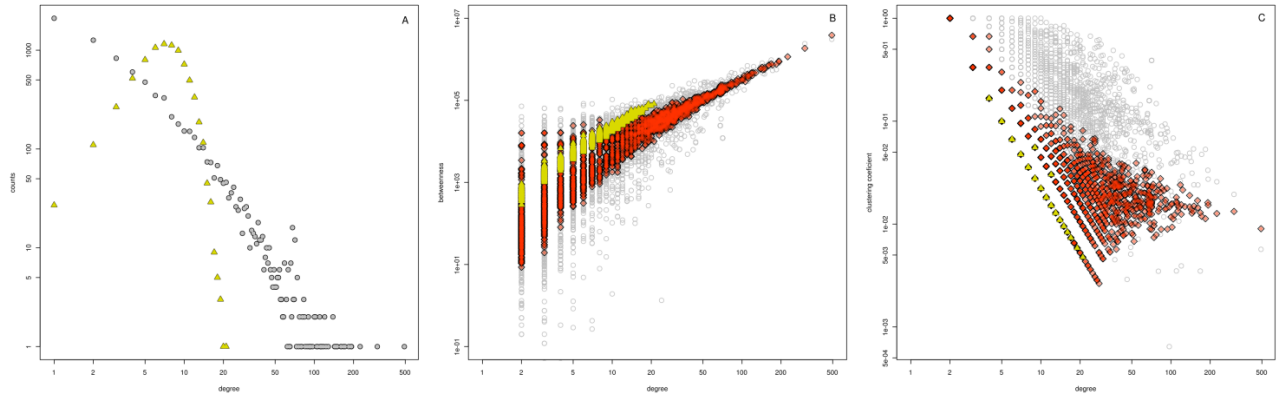

Figure S1 - Topological features of HIPPIE networtk. (A) node degree distribution for HIPPIE network (gray circles), and for the respective Erdos-Renyi random network (yelow triangles). Nodes of the HIPPIE network (gray circles), Rewired Network (red diamonds) and Erdos-Renyi network (yellow triangles) were displayed over the Degree-Betweenness and the Degree-Clustering Coefficient planes in panels (B) and (C) respectively

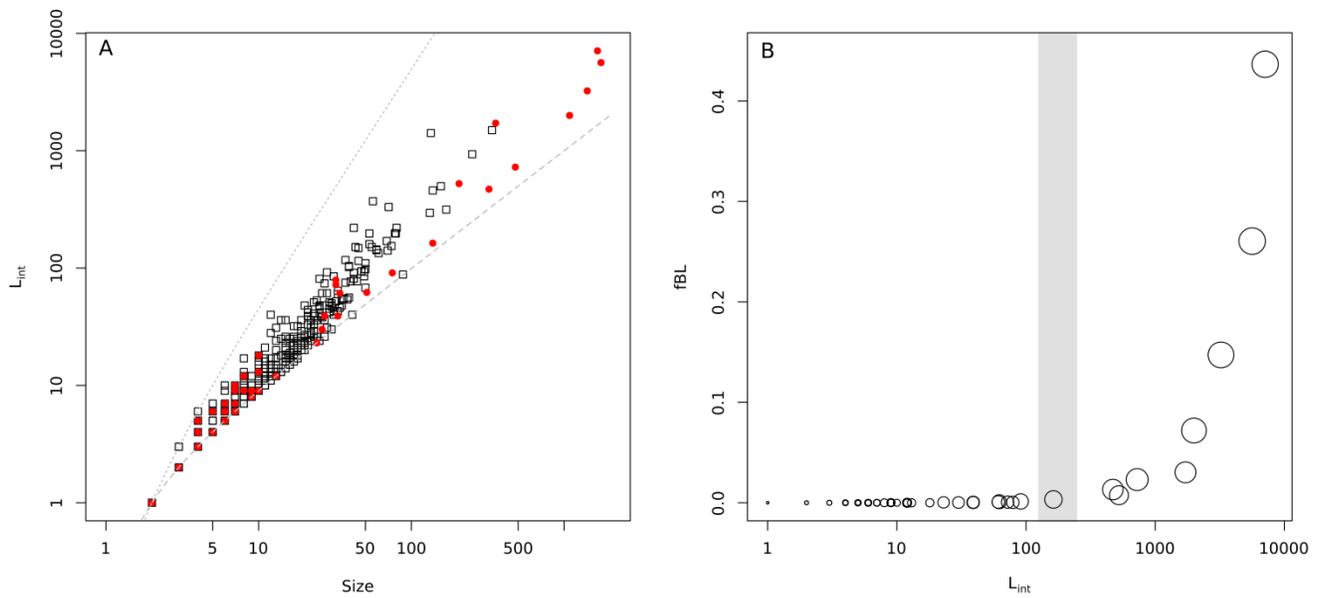

Figure S2 – Modular structure of HIPPIE Network. (a) Number of internal links, $l_{int}$, as a function of cluster sizes. The dotted and dashed lines depict the expected relationships for fully connected cliques and linear structures respectively, and are included for reference purposes. (b) Fraction of internal CNM cluster's links that did not appear as internal links in the infomap modular description (i.e. fraction of broken links, fBL). Circle sizes are proportional to each CNM cluster log-size.

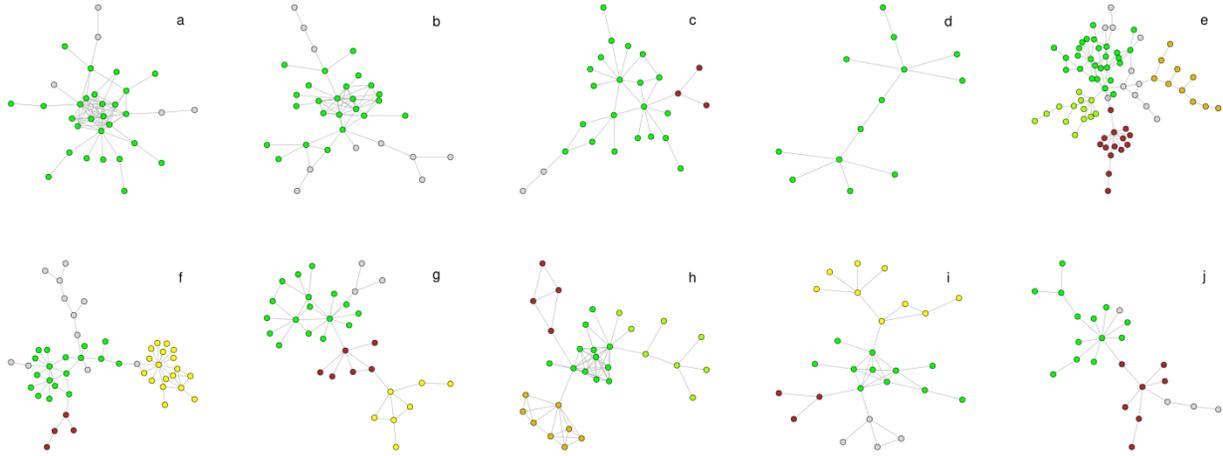

Figure S3 – Examples of module comparison in HN. CNM clusters with low internal-link density and more than ten nodes. infomap communities were depicted using different colors. Gray colored nodes belonged to infomap clusters not-totally included in the displayed CNM structure. Two scenarios can be recognized. For cases (a)-(d) a rather good agreement between the alternative modular descriptions was observed. However, for the cases illustrated in panels (e)-(j) internal structure not resolved by the CNM procedure was indeed highlighted by the infomap prescription.

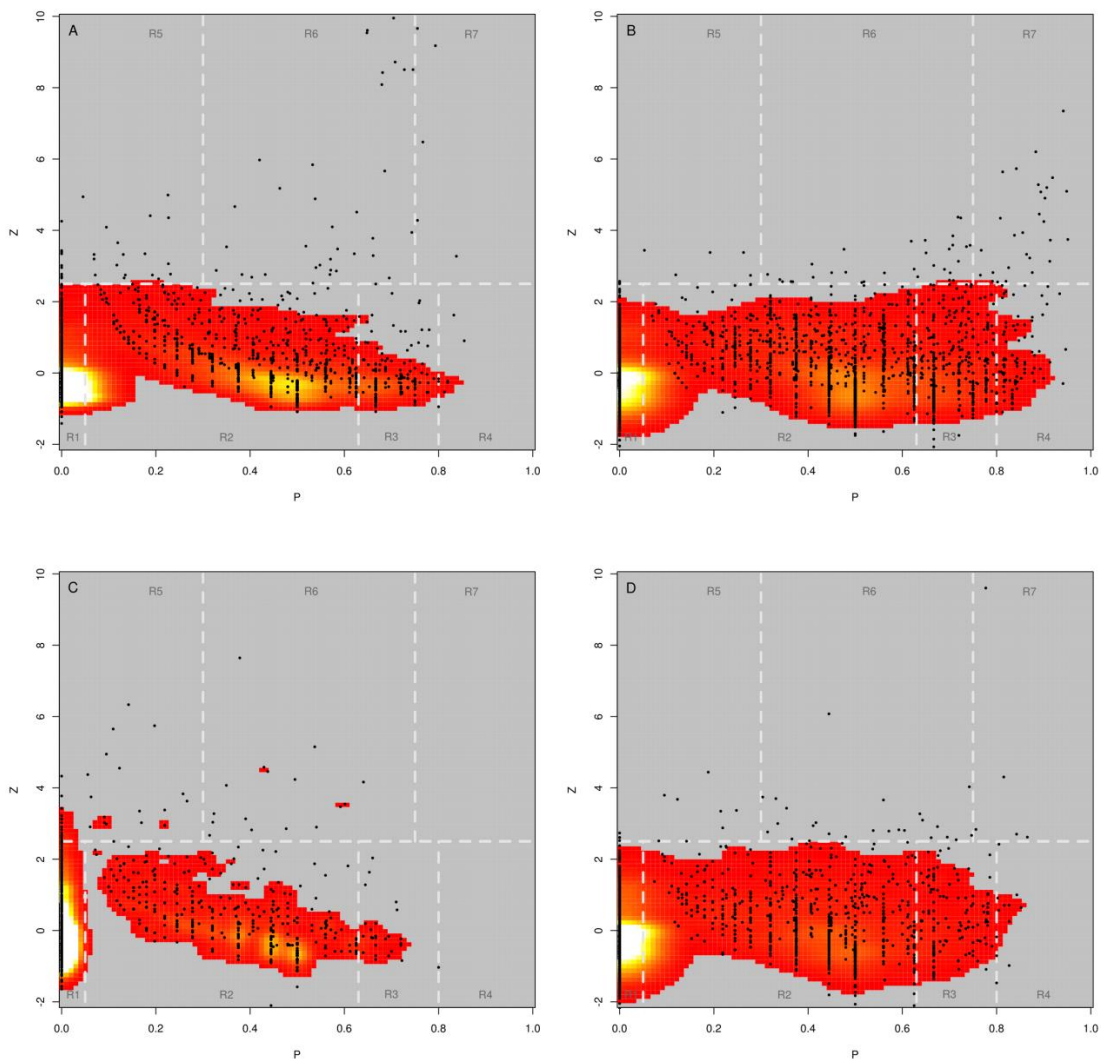

Figure S4 – Cartographic description of already published high confidence PPI networks. Z-P planes for Batada yeast PPI network (panels A-B) and Bertin yeast PPI Network (panels C-D). Left and right panels correspond to *CNM*-based and *infomap*-based cartographical descriptions respectively. An overall increasing behavior in node participation levels can be observed when the *infomap* cluster recognition procedure was considered.

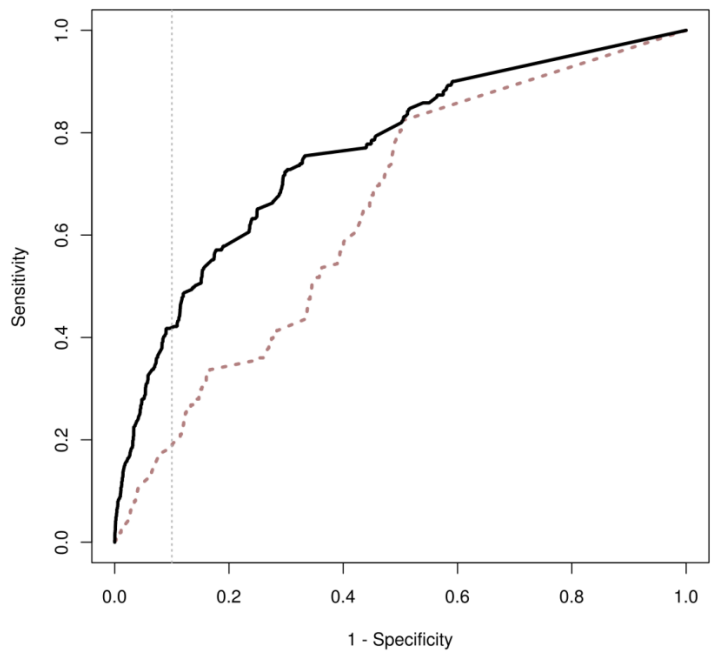

Figure S5 - Participation based analysis of aging related genes. Participation-based ROC curves estimated for ARG genes for *CNM* and *infomap* modular descriptions are shown with dashed brown and continuous black lines respectively. The vertical dotted line represents the 90% specificity level. Statistically significant differences between total AUCs are observed (AUC-IFM =0.76; AUC-CNM=0.65; pvalue<e-16, deLong's test), suggesting that the resolution level provided by *infomap* enhanced the detection of the topological bias displayed by ARG genes toward high-participation levels.

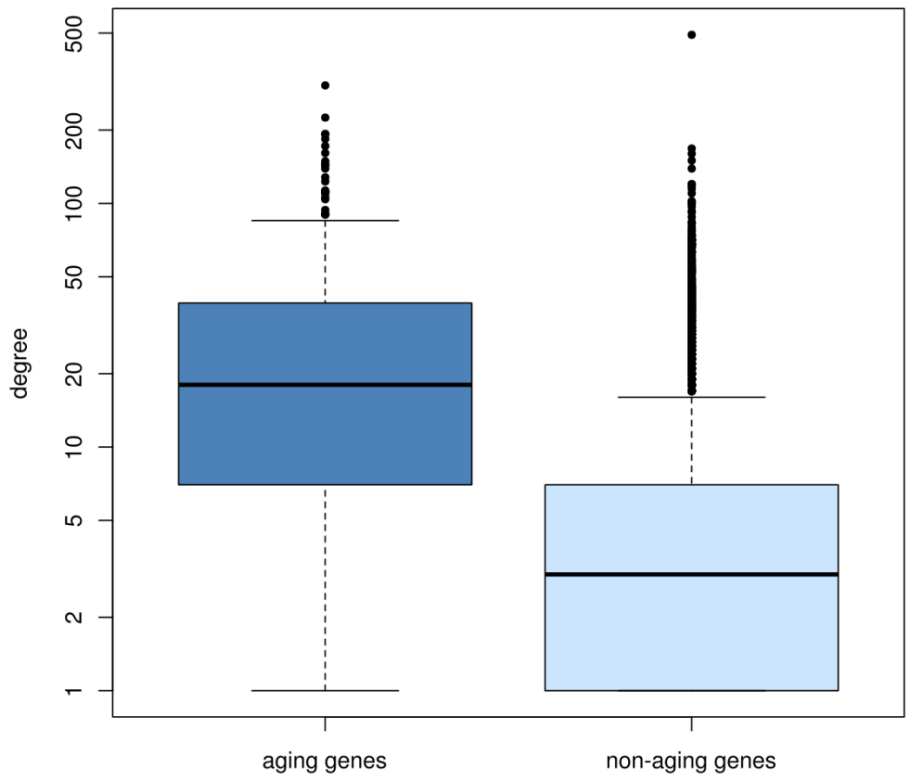

Figure S6 – Evidence of high-degree bias in aging related gene set. The degree distribution of aging genes, and the whole HIPPIE network remaining nodes are shown in the left and right boxplots respectively. Significant differences (pv< e-16, Wilcoxon test) were observed between both degree distributions.

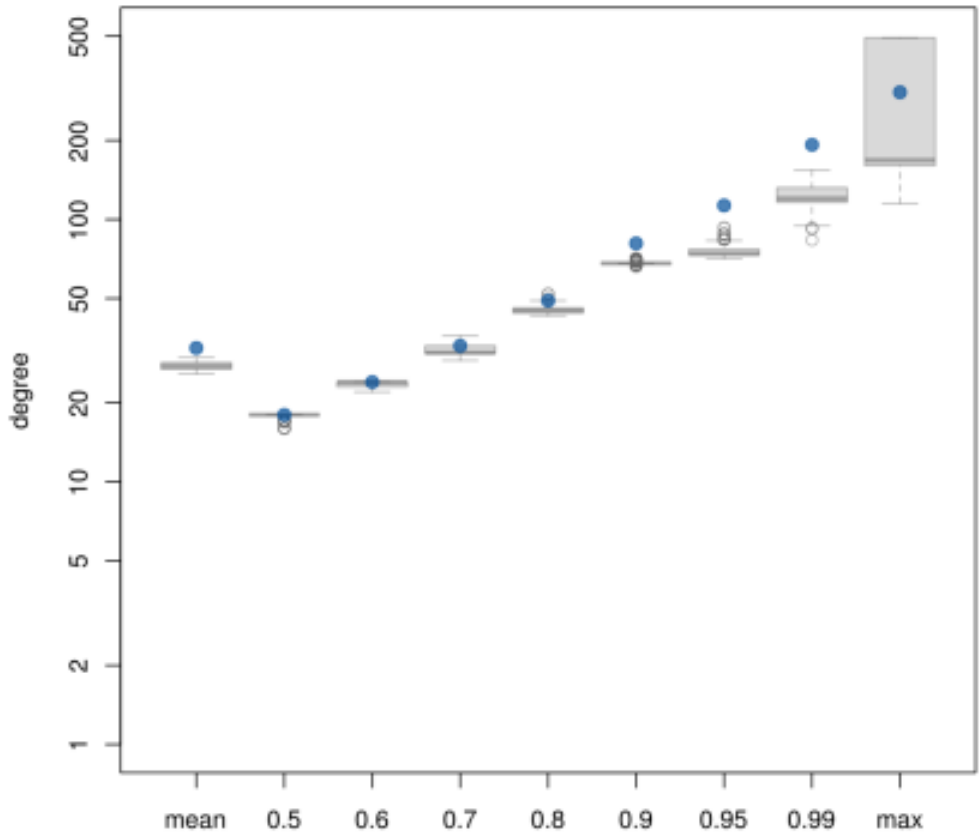

Figure S7 - Control of degree-aware random sampling employed in the bootstrap analysis of aging related gene set. Degree distributions for selected quantiles of 1000 control random realizations are displayed as boxplot. Blue circles depict ARG degree values for the respective quantiles. It can be observed that the top-10% of ARG with highest degree levels, lay outside the inter-quartile levels of their corresponding control random samples.